%% file: main_document.tex
\documentclass[conference]{IEEEtran}

\input{use_packages}
\input{acronyms}
\graphicspath{{./figures/}{./figures/classification/}}
\DeclareGraphicsExtensions{.pdf,.jpg,.png}

\def\BibTeX{{\rm B\kern-.05em{\sc i\kern-.025em b}\kern-.08em
    T\kern-.1667em\lower.7ex\hbox{E}\kern-.125emX}}

\begin{document}
\title{Spatial Consistency Evaluation Based on Massive SIMO Measurements}
\author{
	\IEEEauthorblockN{Sida Dai, Martin Kurras}
	\IEEEauthorblockA{Fraunhofer Heinrich Hertz Institute, Einsteinufer 37, 10587 Berlin, Germany \\
		Email: {martin.kurras@hhi.fraunhofer.de }} 		
}
\maketitle
\begin{abstract}
In this paper, the spatial consistency of wireless massive \acl{SIMO} channels in a cellular small cell scenario is evaluated based on measurements taken in Berlin city. The evaluation is done by computing the similarity of covariance matrices over distance. As similarity measure the \acl{CMD} is used. A classification of the measurement tracks based on the shape of the curves into four different categories is done. 

The results in this paper indicate that spatial consistency is a highly deterministic property in the sense that it depends strongly on the individual environment and not so much on large scale parameters. Therefore, we conclude that spatial consistency is not sufficiently modeled by the current \acl{3GPP} feature.
\end{abstract}
\glsresetall
\section{Introduction} \label{sec:introduction}
\input{sec_introduction}

\section{Spatial Consistency}\label{sec:spatial_cons}
\input{sec_spatial_consistency}
\section{Measurements}\label{sec:measurements}
\input{sec_measurements}

\section{Numerical Results} \label{sec:numerical_results}
\input{sec_numerical_results}
\section{Conclusion} \label{sec:conclusion}


In this paper, we evaluate the spatial consistency feature based on \ac{SIMO} measurements with angular information of the \acp{MPC}. By studying the behavior of \(d_{\mathrm{CMD}}\) over distance we categorize each track into one of the ``\ac{LoS} radial'', ``\ac{LoS} tangential'', ``Far away'' and ``Uncorrelated'' classes. In general, the similarity decreases over distance, however, the speed of decreasing varies form class to class due to individual geometry. Moreover, low similarity is observed in the \ac{NLoS} scenarios, proving it difficult to cluster users in such scenarios. These findings could be beneficial for future \ac{3GPP} proposal in channel model.

This work is a first step to motivate a further and deeper analysis of the measurement data with respect to spatial consistency. As a next step, we will analyze the dependency of the spatial consistency from angular spread and K-factor to better understand in which radio environments massive \ac{MIMO} schemes that utilize similarity in covariance matrices can be applied. This will also include a direct comparison with the \ac{3GPP} spatial consistency feature.
	
\section*{Acknowledgment}\label{sec:acknowledgment}
A part of this work has been performed in the framework of the Horizon 2020 project ONE5G (ICT-760809), funded by the European Union. The authors would like to acknowledge the contributions of their colleagues in the project, although the views expressed in this contribution are those of the authors and do not necessarily represent the project.

Moreover, the authors thank their Fraunhofer \ac{HHI} colleagues for conducting the measurements, which is done by Fabian Undi, Leszek Raschkowski and Stephan Jaeckel. Stephan Jaeckel also did the data post-processing. 
We also thank Boonsarn Pitakdumrongkija\footnote{While Boonsarn was with NEC at the time of the measurements, in the meantime he left NEC.}, Xiao Peng and Masayuki Ariyoshi from NEC Japan for enabling these measurements in the first place.

\bibliographystyle{IEEEtran}
\bibliography{library_spat_cons_measurements}
\end{document}

%% file: use_packages.tex
\usepackage[bookmarks=false]{hyperref}

\usepackage{amsmath}  
\usepackage{amssymb}
\usepackage{amsfonts} 
\usepackage{amsthm}

\usepackage{bm}  

\usepackage{subfigure}

\usepackage{booktabs}

\usepackage{enumitem}

\usepackage{siunitx} 

\usepackage{algorithm} 
\usepackage{algorithmicx}

\usepackage{algpseudocode}

\usepackage{multirow}

\usepackage{longtable}

\usepackage{dsfont}

\usepackage{xcolor} 
\definecolor{hyp}{rgb}{0,0,1} 
\definecolor{GrayA}{gray}{0.95}
\definecolor{GrayB}{gray}{0.90}
\definecolor{GrayC}{gray}{0.85}		

\usepackage{colortbl} 
\newcolumntype{L}[1]{>{\raggedright\let\newline\\\arraybackslash\hspace{0pt}}m{#1}}
\newcolumntype{C}[1]{>{\centering\let\newline\\\arraybackslash\hspace{0pt}}m{#1}}
\newcolumntype{R}[1]{>{\raggedleft\let\newline\\\arraybackslash\hspace{0pt}}m{#1}}

\usepackage{graphicx}
\graphicspath{{figures/}}
\DeclareGraphicsExtensions{.pdf,.jpg,.png,.pdf_tex,.eps,.gif}

\usepackage{epstopdf} 

\usepackage{placeins}

\usepackage{scrhack} 

\usepackage[acronym,shortcuts,nomain]{glossaries}
\newglossarystyle{super_ext} 
{
		\setglossarystyle{super}

}

\newcommand{\acro}[2]{\newacronym{#1}{#1}{#2}}

\setglossarystyle{super_ext}

\usepackage[capitalise]{cleveref}

%% file: acronyms.tex
\acro{2D}{two dimensional}
\acro{2G}{second generation}
\acro{3D}{three dimensional}
\acro{3G}{third generation}
\acro{3GPP}{3rd Generation Partnership Project}
\acro{4G}{fourth generation}
\acro{5G}{fifth generation}
\acro{ACF}{auto correlation function}
\acro{ACK}{acknowledgment}
\acro{ADC}{analog to digital converter}
\acro{AGNSS}{assisted \acl{GNSS}}
\acro{AMC}{adaptive modulation and doding}
\acro{AoA}{angle of arrival}
\acro{AoD}{angle of departure}
\acro{ARQ}{automatic repeat request}
\acro{ARTES}{advanced research in telecommunications systems}
\acro{ASSQ}{adaptive search space quantization}
\acro{AVI}{actual value interface}
\acro{AWGN}{additive white Gaussian noise}
\acro{BBU}{base band unit}
\acro{BC}{broadcast channel}
\acro{BD}{block diagonalization}
\acro{BER}{bit error rate}
\acro{BLER}{block error rate}
\acro{BMWi}{Federal Ministry for Economic Affairs and Energy}
\acro{BPCU}{bits per channel use}
\acro{BPSK}{binary phase-shift keying}
\acro{BS}{base station}
\acro{CAPEX}{capital expenditure}

\acro{CCDF}{complementary cumulative distribution function}
\acro{CCI}{co-channel interference}
\acro{CDF}{cumulative distribution function}
\acro{CDMA}{code-division multiple access}
\acro{CFMAI}{cumulative feedback map with aged information}
\acro{CIR}{channel impulse response}
\acro{CMD}{correlation matrix distance}
\acro{CoMP}{coordinated multi-point}
\acro{COST}{European Cooperation in Science and Technology}
\acro{CP}{cyclic prefix}
\acro{CQI}{channel quality indicator}
\acro{CRB}{Cramer Rao bound}
\acro{CRLB}{Cramer Rao lower bound}
\acro{CRC}{cyclic redundancy check}
\acro{CRONOS}{cellular radio network simulator}
\acro{CRS}{cell-specific reference signal}
\acro{CSI}{channel state information}
\acro{CSIR}{\ac{CSI} at the receiver}
\acro{CSIT}{\ac{CSI} at the transmitter}
\acro{CS}{compressed sensing}
\acro{D2D}{device-to-device}
\acro{DAS}{distributed antenna system}
\acro{dB}{decibel}
\acro{DBSCAN}{density based clustering on application with noise}
\acro{DCM}{database correlation method} 
\acro{DFT}{discrete Fourier transform}
\acro{DG-JDE}{device grouping for joint \ac{DoA} estimation}
\acro{DMRS}{demodulation reference signals} 
\acro{DoA}{direction of arrival}
\acro{DoF}{degree of freedom}
\acro{DoP}{dilution of precision}
\acro{DRS}{demodulation reference signals}
\acro{DPC}{dirty paper coding}
\acro{DS}{delay spread}
\acro{DSSS}{direct sequence spread spectrum}
\acro{EBF}{eigen-beamforming}
\acro{ECE}{effective channel estimation}
\acro{EESM}{exponential effective \ac{SINR} mapping}
\acro{eNB}{enhanced node b}
\acro{EPS}{equal power per stream}
\acro{ESA}{European space agency}
\acro{ESPRIT}{estimation of signal parameter via rotational invariance technique}
\acro{FDD}{frequency division duplex}
\acro{FEC}{forward error correction}
\acro{FFT}{fast Fourier transform}
\acro{FFR}{fractional frequency reuse}
\acro{FI}{feedback interval}
\acro{FSK}{frequency shift keying}
\acro{GDoP}{geometry dilution of precision}
\acro{GoB}{grid of beams}
\acro{GNSS}{global navigation satellite system}
\acro{GPS}{global positioning system}
\acro{GPRS}{general packet radio service}
\acro{GSB}{\acs{GoB} subsets block}
\acro{GSCM}{geometry-based stochastic channel model}
\acro{GSM}{global system for mobile communications}

\acro{HARQ}{hybrid \ac{ARQ}}
\acro{HetNet}{heterogeneous network}
\acro{HHI}{Heinrich Hertz Institute}
\acro{HPBW}{half-power beam-width}
\acro{HSDPA}{high speed downlink packet access}
\acro{ICT}{information and communications technology}
\acro{ICIC}{inter-cell interference coordination}
\acro{IDFT}{inverse discrete Fourier transform}
\acro{IEEE}{institute of electrical and electronics engineers}
\acro{IFFT}{inverse \ac{FFT}}
\acro{i.i.d.}{independent identically distributed}
\acro{IoT}{internet of things}
\acro{IRC}{interference rejection combining}
\acro{ISD}{inter-site distance}
\acro{ITU}{International Telecommunications Union}
\acro{ITU-R}{\acs{ITU} radio communication sector}
\acro{JGP}{joint group processing}
\acro{JSDM}{joint spatial division and multiplexing}
\acro{KF}{Rician K-factor}
\acro{KPI}{key performance indicator}
\acro{L2S}{link to system}
\acro{LDPC}{low density parity check}
\acro{LGPL}{Lesser General Public License}
\acro{LHS}{left-hand side}
\acro{LoS}{line of sight}
\acro{LSAS}{large scale antenna system}
\acro{LSP}{large scale parameter}
\acro{LTE}{Long Term Evolution}
\acro{LTE-A}{\ac{LTE}-Advanced}
\acro{LTTS}{long-term time sample}
\acro{MCS}{modulation and coding scheme}
\acro{MDE}{multiple device estimation}
\acro{MET}{multiple user eigenmode transmission}
\acro{METIS}{Mobile and wireless communications Enablers for the Twenty-twenty Information Society}
\acro{MIESM}{mutual information effective \ac{SINR} mapping}
\acro{MIMO}{multiple-input multiple-output}
\acro{MISO}{multiple-output single-input}
\acro{ML}{maximum likelihood}
\acro{MSLR}{minimum side lobe ration}
\acro{MMIB}{mean mutual information per bit}
\acro{MMSE}{minimum mean square error}
\acro{MPC}{multi-path component}
\acro{MRC}{maximum ratio combining}
\acro{MRT}{maximum ratio transmission}
\acro{MS}{mobile station}
\acro{MSE}{mean square error}
\acro{MU}{multiple-user}
\acro{MUSIC}{multiple signal classification}
\acro{MVDR}{minimum variance distortionless response}
\acro{NLoS} {non-line of sight}
\acro{NR}{new radio}
\acro{OAM}{operation and maintenance}
\acro{OCEAN} {operating cellular large scale \ac{MIMO} for energy efficient networks}
\acro{OFDM}{orthogonal frequency division multiplexing}
\acro{OFDMA}{orthogonal frequency division multiple access}
\acro{OPEX}{operational expenditure}
\acro{OPTICS}{ordering points to identify the clustering structure}
\acro{OTDoA}{observed time difference of arrival}
\acro{PA}{power amplifier}
\acro{PAM}{pulse amplitude modulation}
\acro{PAPC}{per antenna power constraint}
\acro{PAPR}{peak-to-average power ratio}
\acro{PBZF}{projection based zero forcing}
\acro{PCM}{pilot correlation method}
\acro{PDF}{probability distribution function}
\acro{PDP}{power delay profile}
\acro{PER}{packet error rate}
\acro{PhD}{Doctor of Philosophy}
\acro{PGP}{per group processing}
\acro{PHY}{physical}
\acro{PMI}{precoding matrix indicator}
\acro{PN}{pseudo noise}
\acro{PO}{pilot overhead}
\acro{PRB}{physical resource block}
\acro{PUSCH}{physical uplink shared channel}
\acro{QuaDRiGa}{quasi-deterministic radio channel generator}
\acro{QAM}{quadrature amplitude modulation}
\acro{QoS}{quality of service}
\acro{QPSK}{quadrature phase shift keying}
\acro{RADAR}{radio detection and ranging}
\acro{RAN1}{radio access network one}
\acro{RB}{resource block}
\acro{RE}{resource element}
\acro{RF}{radio frequency}
\acro{RHS}{right-hand side}
\acro{RMS}{root mean square}
\acro{RRM}{radio resource management}
\acro{RRU}{remote radio unit}
\acro{RS}{reference signal}
\acro{RSRP}{\ac{RS} receive power}
\acro{RSS}{receive signal strength}
\acro{RSSI}{received signal strength indicator}
\acro{RTT}{round trip time}
\acro{RVQ}{random \ac{VQ}}
\acro{RX}{receiver}
\acro{RZF}{regularized \ac{ZF}}
\acro{SC}{single carrier}
\acro{SC-FDMA}{\ac{SC} frequency domain multiple access}
\acro{SCME}{spatial channel model extended}
\acro{SDE}{single device estimation}
\acro{SDMA}{spatial division multiple access}
\acro{SE}{spectral efficiency}
\acro{SIC}{successive interference cancellation}
\acro{SIMO}{single-input multiple-output}
\acro{SINR}{signal to interference and noise ratio}
\acro{SIR}{signal to interference ratio}
\acro{SISO}{single-input single-output}
\acro{SLNR}{signal to leakage and noise ratio}
\acro{SMUX}{spatial multiplexing}
\acro{SMS}{short message service}
\acro{SNR}{signal to noise ratio}
\acro{SOS}{sum of sinusoids}
\acro{SRS}{sounding reference signals}
\acro{SSF}{small-scale fading}
\acro{STTS}{short-term time sample}
\acro{SU}{single-user}
\acro{SUS}{semi-orthogonal user selection}
\acro{SVD}{singular value decomposition}
\acro{TDD}{time division duplex}
\acro{TDMA}{time division multiple access}
\acro{TDoA}{time difference of arrival}
\acro{ToA}{time of arrival}
\acro{ToF}{time of flight}
\acro{TP}{throughput}
\acro{TRX}{transceiver}
\acro{TTI}{transmission time interval}
\acro{TUB}{Technische Universit{\"a}t Berlin}
\acro{TX}{transmitter}

\acro{UAV}{unmanned areal vehicle}
\acro{UCA}{uniform cylindrical array}
\acro{UE}{user equipment}
\acro{UL}{uplink}
\acro{ULA}{uniform linear array}
\acro{UMa}{urban macro}
\acro{UMTS}{universal mobile telecommunications system}
\acro{UNP}{user non-aware precoding}
\acro{UPA}{uniform planar array}
\acro{URA}{uniform rectangular array}
\acro{USA}{uniform square array}
\acro{UWB}{ultra wideband}
\acro{VQ}{vector quantization}
\acro{WCDMA}{wideband code division multiple access}
\acro{WINNER}{wireless world initiative new radio}
\acro{WLAN}{wireless local area network}
\acro{WLS}{weighted least square}
\acro{w.r.t.}{with respect to}
\acro{XPR}{cross-polar ratio}

\acro{ZC}{Zadoff-Chu}
\acro{ZF}{zero forcing}

%% file: sec_introduction.tex
In wireless communications research, simulations play an important role. They enable early and accurate evaluation or comparison of new techniques, for example in \ac{3GPP} standardization \cite{3GPP17-38901} or in most of the recent research papers. An essential part of each wireless transmission simulation is the underlying channel and the modeling of it \cite{MJY10}. There are various channel models but it is out of the scope of this paper to provide an overview; the interested reader is referred to \cite{Jae17}. Assumptions taken for the modeling of the channel directly impact or limit simulation results. For example, while the assumption of independent fading between users may hold in a multiple-user \ac{MIMO} transmission, this assumption cannot be used to evaluate techniques utilizing channel correlation. One example of such a scheme is \ac{JSDM} \cite{ANAC13}, where users are clustered into groups with similar channel properties prior to data-transmission. 

It is intuitive that users close to each other experience similar wireless channels to the same transmitter. The feature that models such behavior, called \textit{spatial consistency}, was first introduced in \cite{3GPP14-TR25996} and further discussed in \cite{3GPP17-38901} and researchers have been focusing on this topic ever since. 
In \cite{KDJT18} Kurras et. al. provide an evaluation of the \ac{3GPP} spatial consistency feature implemented in the open source channel model \ac{QuaDRiGa} \cite{JRB+17}, taking into account angular distance, chordal distance and the \ac{CMD}. 
In \cite{WSH+16} measurements at \SI{73}{\giga\hertz} are conducted with a single horn antenna at both ends, where the transmit antennas are fixed and the receive antennas sweep with a \SI{5}{\degree} resolution.
With such a measurement setup the angular spread of \ac{MPC} can not sufficiently be measured. 
Additionally, with a wavelength of \(\approx\SI{0.41}{\milli \meter}\) and a given measurement density of \SI{0.2}{\meter} which is approximately a resolution of 50 wavelengths, the changes in the phases of the multi-path components can not be captured accurately.
\cite{GLH+17} provides a purely simulation based study where the spatial consistency of dominant \acp{MPC} in a high-speed train scenario is evaluated comparing the \ac{3GPP} channel model with ray-tracing. In \cite{BPC+17}, the spatial consistency in vehicles is compared between the \(3-\SI{11}{\giga\hertz}\) and \(55-\SI{65}{\giga\hertz}\) bands. However, the evaluation only focuses on power-delay, without taking any angular information into account. 

In this paper, we use \ac{SIMO} measurements including angular information of the \acp{MPC}\cite{RJU+16} to provide an evaluation of the spatial consistency feature. The angular information is obtained with the help of a cylindrical antenna array at the receive \ac{BS}. With a measurement center frequency of \SI{3.675}{\giga\hertz} in a small-cell like deployment, the result shown in this paper can directly be considered for the upcoming cellular \ac{5G} systems\footnote{\ac{MIMO} simulation assumptions in \ac{3GPP} consider \SI{4}{\giga\hertz} center frequency, see Section 7.1.6 in \cite{3GPP17-38802} .}. 

In the remainder of this paper, we provide details on the \ac{CMD} evaluation metric for spatial consistency in \cref{sec:spatial_cons}. In \cref{sec:measurements}, details of the underlying measurement campaign are given, which are used for results in \cref{sec:numerical_results}. Finally, \cref{sec:conclusion} summarizes our findings and concludes this paper. 

%% file: sec_spatial_consistency.tex
Previous \acp{GSCM} suffer from the lack of realistic correlation in the \ac{SSF}, i.e. even though the \acp{LSP} are always spatially consistent, the models fail to provide correlated \ac{SSF}, governed by the position of the scattering clusters.
The \ac{3GPP} proposed a new channel model for the upcoming \ac{5G} of wireless communications in \cite{3GPP17-38901}. This model solves the drawbacks in previous \acp{GSCM} and a part of it is the introduction of the so called \textit{spatial consistency} feature. 
With spatial consistency, two closely located users will not only have similar \acp{LSP}, they will also observe similar angles of received \acp{MPC}.


In order to evaluate the spatial consistency feature, authors in \cite{KDJT18} compared different performance metrics, among which the covariance matrix based ones have provided reasonable and representative results.
The covariance matrix is assumed to be slow time-varying and is therefore a suitable \ac{KPI} for evaluating the spatial correlation between users \cite{ANAC13, KDJT18}. Thanks to this property, lots of studies on user clustering are established on covariance matrix based similarity measures (e.g. chordal distance \cite{GVL12, ANAC13} or \ac{CMD} \cite{MHA+17}). Therefore, this paper focuses on the covariance matrix based \ac{CMD} metric for evaluation of the spatial consistency and details on how the covariance matrix is obtained are provided next.
\subsection{Covariance Matrix}\label{sec:cov_mtrx}
The covariance matrix is obtained by averaging the channel coefficients over the time duration $\tau$ and the number of \acl{OFDM} subcarriers $N$. Channel coefficients between the transmitter and the receiver at time $t$ on the $n$-th subcarrier are denoted as \(\mathbf{H}_{t,n} \in \mathbb{C}^{n_r \times n_t}\), where \(n_r\) and \(n_t\) are the number of antennas at the receiver and the transmitter, respectively. 
The covariance matrix at the receiver side $\mathbf{R}\in\mathbb{C}^{n_r\times n_r}$ is often defined as 
\begin{equation}
	\mathbf{R}^{(\mathrm{Lit})} = \mathbb{E}\left[\mathbf{H}_{t,n}\mathbf{H}^\mathrm{H}_{t,n}\right],
\label{eq:cov_literature}
\end{equation}
without further explanation on how exactly the covariance matrix can be obtained \cite{ANAC13,MHA+17}, because the covariance matrix is directly generated in the simulations and Gaussian noise is added for individual channel realizations. In \cref{eq:cov_literature}, $\mathbb{E}[\cdot]$ denotes the expectation value. However, in our case the covariance matrix has to be obtained from discrete measurements and is calculated as
\begin{equation}
	\mathbf{R}^{(\mathrm{Meas})}=\frac{1}{\tau N}\sum_{t=1}^{\tau}\sum_{n=1}^N\mathbf{H}_{t,n}\mathbf{H}^\mathrm{H}_{t,n}.
	\label{eq:cov_measurements}
\end{equation}
It can be seen from \cref{eq:cov_measurements} that the covariance matrix depends on the selection of the averaging time \(\tau\) and the averaging bandwidth represented by \(N\).

Based on the covariance matrix according to \cref{eq:cov_measurements}, details on the \ac{CMD} to study the spatial correlation between two user is given next.
\subsection{Correlation Matrix Distance}\label{sec:cmd}
The \ac{CMD} was proposed in \cite{GVL12} and served as a novel measure to track the changes in spatial structure of non-stationary \ac{MIMO} channels.
Results in \cite{KDJT18} have shown a strong correlation between the physical distance and the \ac{CMD}.
Given the covariance matrices of two users ($\mathbf{R}_1$,$\mathbf{R}_2$), the similarity measure based on \ac{CMD}, according to \cite{GVL12}, can be obtained by
\begin{equation}
	\label{eq:cmd}
	d_\mathrm{CMD}\left(\mathbf{R}_1,\mathbf{R}_2\right) =\frac{\mathrm{Tr}(\mathbf{R}_1^\mathrm{H}\mathbf{R}_2)}{\|\mathbf{R}_1\|_\mathrm{F}\cdot\|\mathbf{R}_2\|_\mathrm{F}} \\
\end{equation}
where \(\mathrm{Tr}(\cdot)\) denotes the ``trace'' operator. 
The \ac{CMD} based similarity measure is a normalized metric which is upper-bounded by $1$ in the case of $\mathbf{R}_1$ and $\mathbf{R}_2$ being collinear, and lower-bounded by $0$ in the case of $\mathbf{R}_1$ and $\mathbf{R}_2$ being orthogonal.

%% file: sec_measurements.tex
\begin{figure*}
	\includegraphics[width=0.95\textwidth]{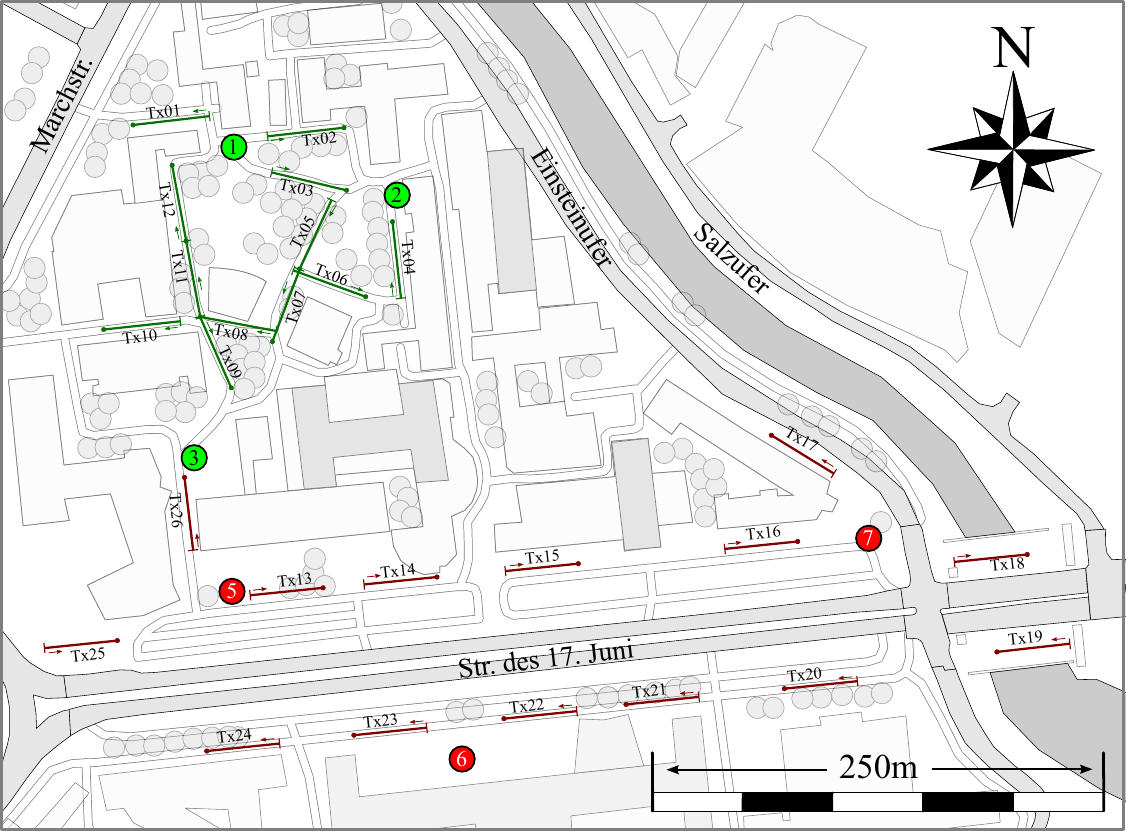}
	\caption{Map of measurement locations in Berlin. Circles with numbers represent \ac{BS} locations and ``Tx01'' - ``Tx26'' indicate measurement tracks.}
	\label{fig:meas_map}
	\vspace{-0.3cm}
\end{figure*}

The massive \ac{SIMO} small cell measurements have been taken in Berlin City close to Fraunhofer \ac{HHI} main building. \cref{fig:meas_map} shows a map of the measurement area, where circles with numbers 1 to 3 and 5 to 7 represent \ac{BS} locations and the colored lines with text ``Tx01'' to ``Tx26'' next to it denote the measurement tracks. The strokes at one end together with small arrows indicate the starting point of the tracks and circles at the other end represent the end point. At each \ac{BS} location two different \ac{BS} heights of \SI{3}{\meter} and \SI{6}{\meter} have been measured. The \ac{BS} locations 1 to 3 are associated with tracks 1 to 12 and belong to the ``Campus'' scenario. The remaining \ac{BS} locations 5 to 7 with measured tracks 13 to 26 belong to the ``Open Street'' scenario. It can be seen from \cref{fig:meas_map} that the tracks cover \ac{LoS} and \ac{NLoS} situations in both scenarios.

\begin{figure}
	\centering
	\subfigure[Side view]{
			\includegraphics[width=0.46\linewidth]{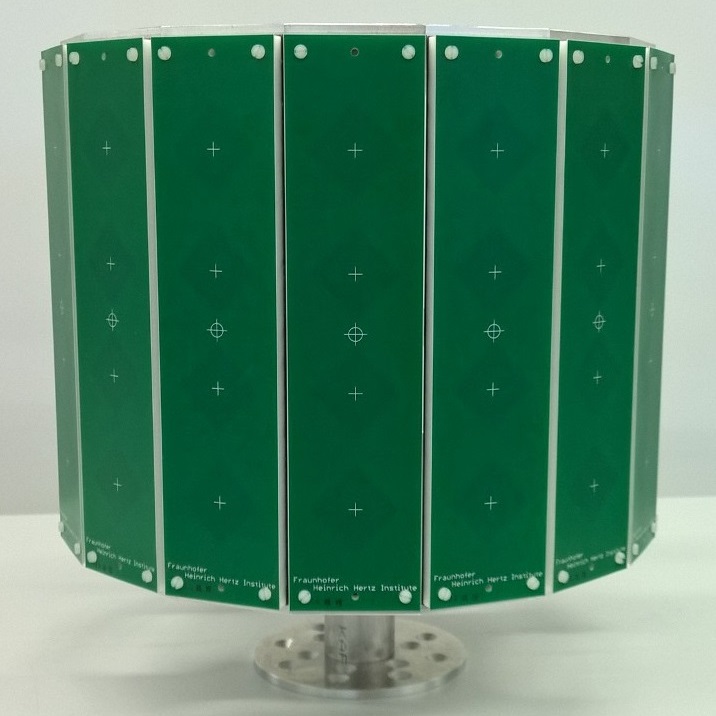}
		\label{fig:antenna_side}
		}
	\subfigure[Top View]{
		\includegraphics[width=0.46\linewidth]{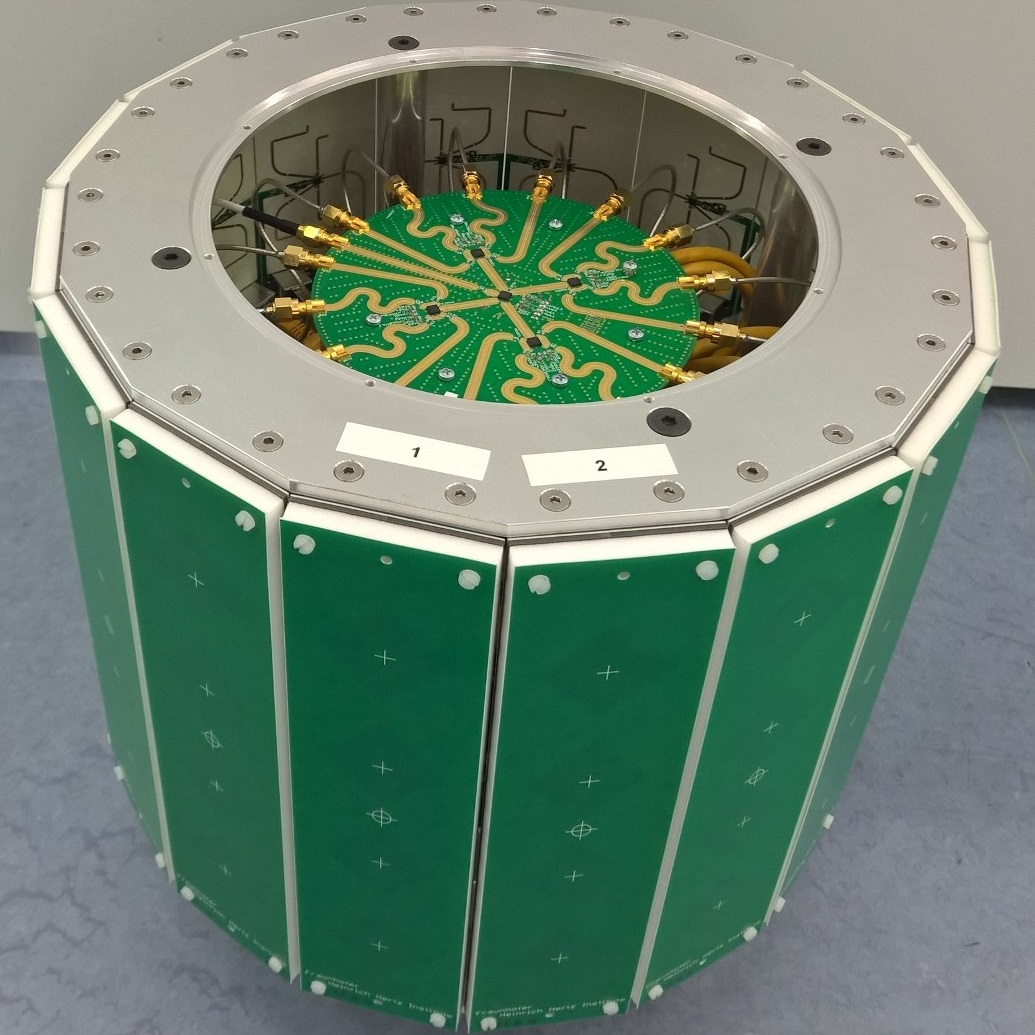}
		\label{fig:antenna_top}
		}
	\caption{Manufactured switched cylindrical array antenna with 128 elements and the switched measurement timing \cite{RJU+16}.}
	\label{fig:antenna_pic}
\end{figure}

For the \ac{BS} we used a \ac{UCA} with 16 columns and 4 stacked dual-polarized patch antennas resulting in 128 antenna elements in total, shown in \cref{fig:antenna_pic}. Inside the \ac{UCA} a ``128 to 1'' \ac{RF} self-developed switch was used making it possible to measure all antenna with the same \ac{RF} receiver chain sequentially. For transmit antenna at the mobile device an omni-directional vertically polarized single antenna from Huber+Suhner model ``1399.17.0111'' was used.

\begin{figure}
	\centering
	\subfigure[6 m]{
			\includegraphics[height=0.62\linewidth]{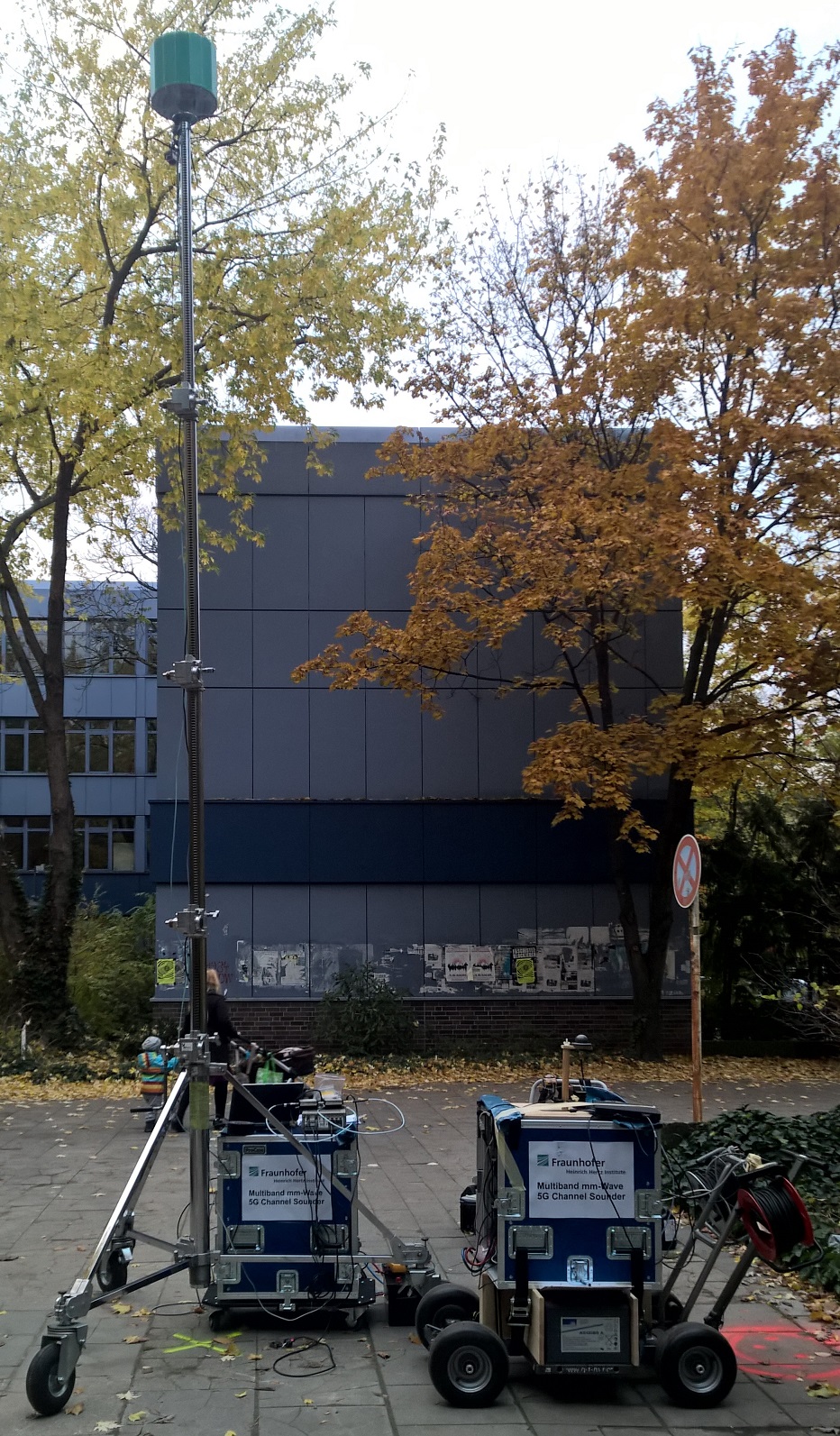}
		\label{fig:tripod_high}
		}
	\subfigure[3 m]{
		\includegraphics[height=0.62\linewidth]{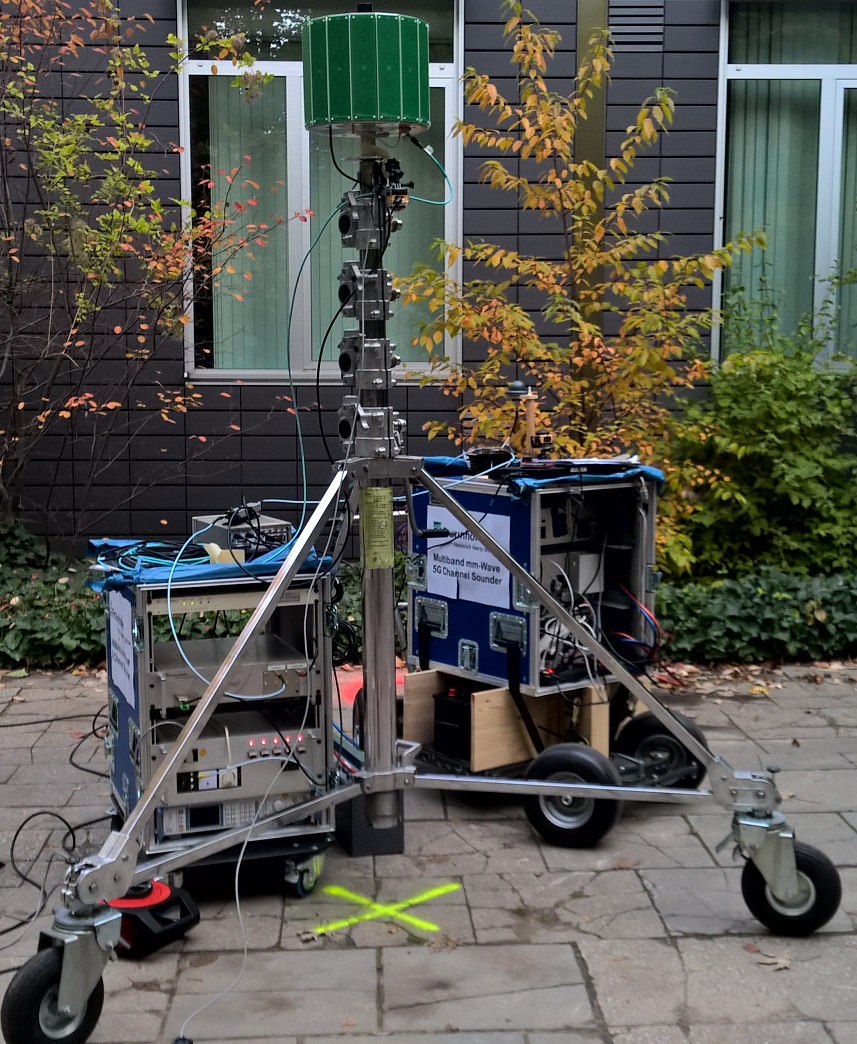}
		\label{fig:tripod_low}
		}
	\subfigure[Transmitter moving on cart]{
		\includegraphics[width=0.92\linewidth]{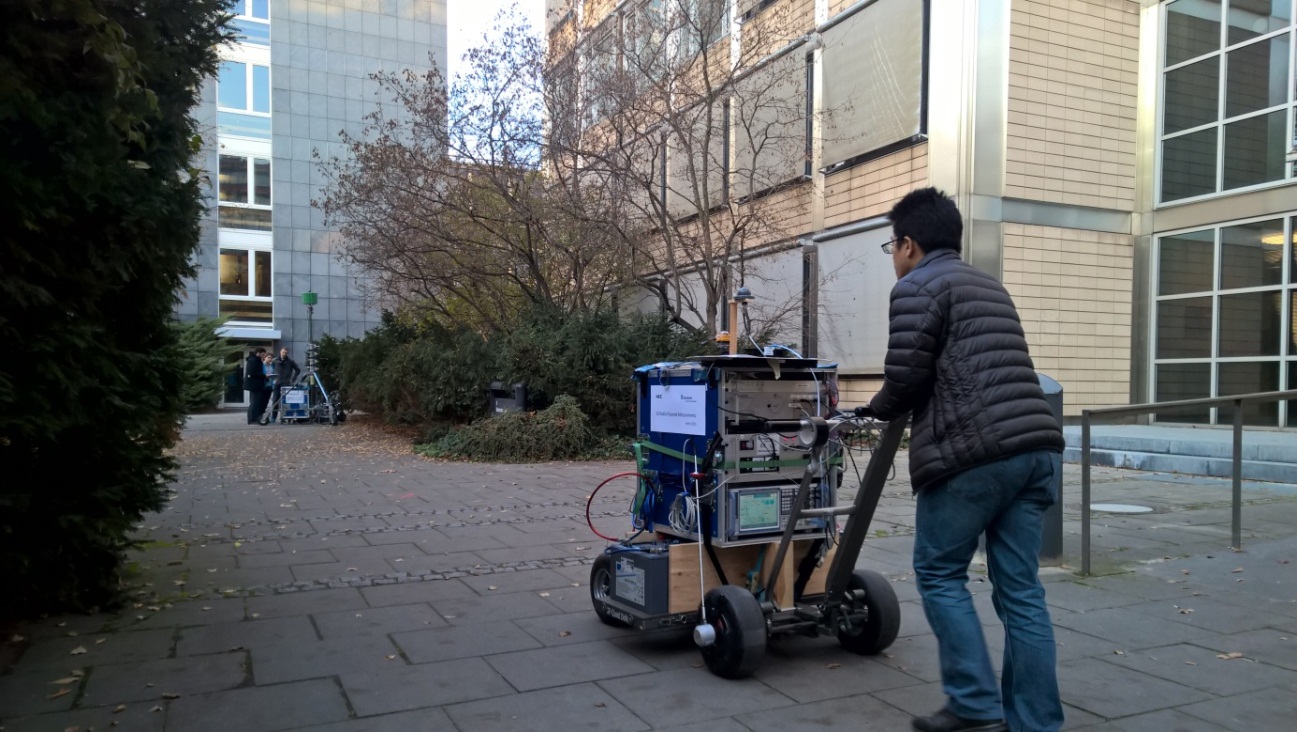}
		\label{fig:timing}
		}

	\caption{Measurement equipment. The cylindrical receive antenna is mounted on a tripod at (a) 6～m and (b) 3~m height. (c) The transmitter is mounted on a dolly and moving at a speed of \(\approx \SI{1.8}{\kilo\meter/\hour}\). \cite{RJU+16}}
	\label{fig:equipment}
\end{figure}

\begin{figure}
    \includegraphics[width=\columnwidth]{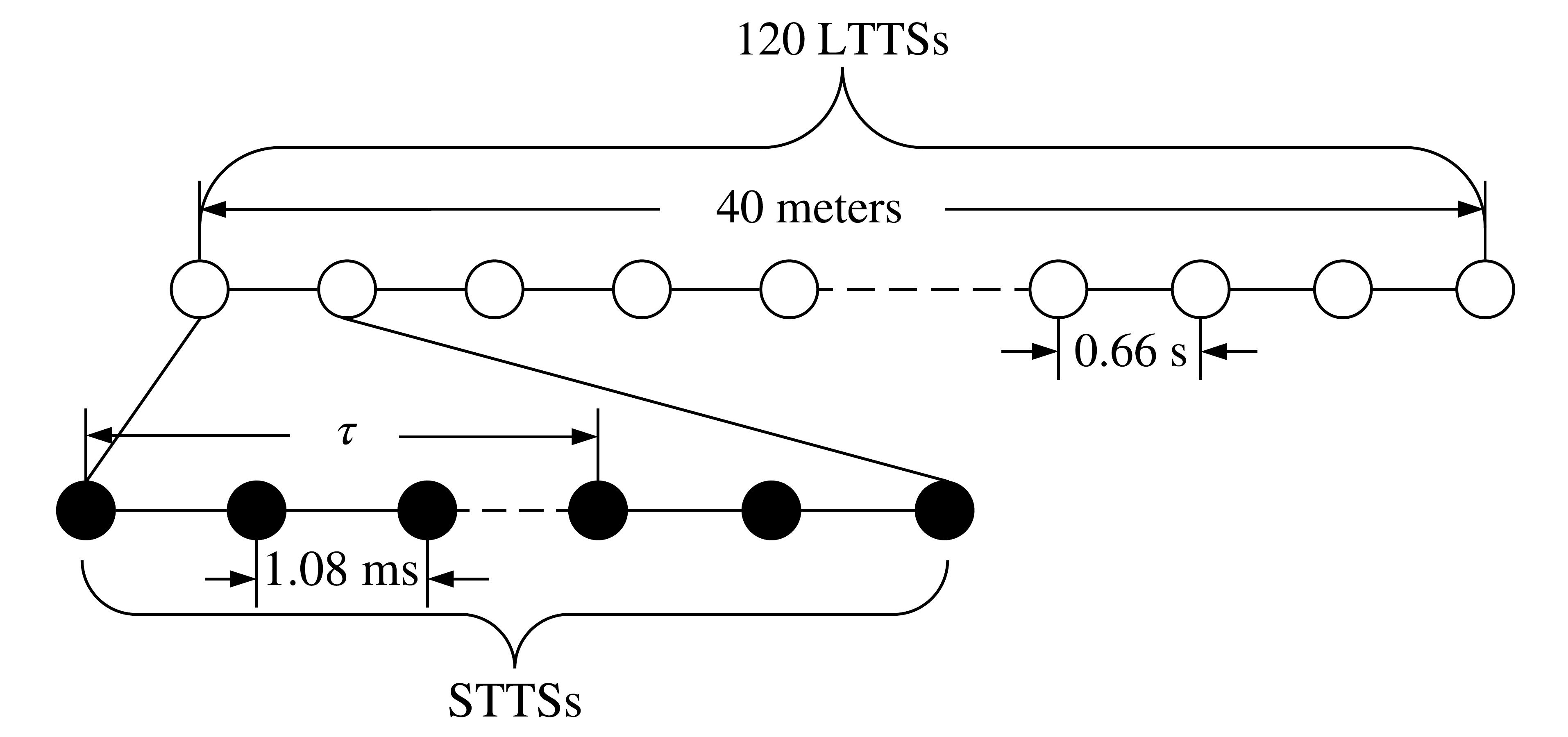}
    \caption{Illustration of a sample track. The first $\tau$ \acp{STTS} are used for covariance matrix calculation.}
    \label{fig:track_view}
\end{figure}

The complete measurement setup is shown in \cref{fig:equipment}. During the measurements the transmit antenna was moved along the \SI{40}{\meter} tracks at a height of \SI{1.5}{\meter} with a constant speed of \(\approx \SI{1.8}{\kilo\meter/\hour}\). The \ac{BS} antenna was receiving Chadoff-Zho like sounding sequences of 1024 sample-length with a duration of \SI{4.1}{\micro\second}. Switching all 128 antenna elements at the \ac{BS}, called a complete \ac{SIMO} short-term time sample (STTS), was measured every \SI{1.08}{\milli\s}. This includes already some real-time averaging. The signal bandwidth was \SI{250}{\mega\hertz} with a center frequency of \SI{3.675}{\giga\hertz}. Transmitter and receiver have been synchronized by a self-developed Rubidium clock. A \ac{LTTS} consists of numerous \acp{STTS}, where \(\tau\) out of them are used for covariance matrix averaging, and has a time interval of \SI{0.66}{\s}, see \cref{fig:track_view}. Each track is divided into 120 \acp{LTTS}. Further details of the measurement campaign are given by Raschkowski et. al. in \cite{RJU+16} and by Pitakdumrongkija et. al. in \cite{PAR+16}.

%% file: sec_numerical_results.tex
\begin{table}
	\centering
	\begin{tabular}{m{3cm}|C{3cm}}
		Parameter 												& Value \\ \hline \hline
		Measurement direction							& Uplink \\ \hline
		Center frequency 									& \SI{3.675}{\giga\hertz} \\ \hline
		Measured scenarios								& Open Square, Campus\\ \hline
		Propagation scenarios							& LoS, NLoS \\ \hline
		Number \acs{BS}	antennas \(n_r\)         	& 128 \\ \hline
		Distribution \acs{BS} antennas		& \ac{UCA}, 16 columns, 4 rows, 2 polarizations, see Fig.~2 in \cite{RJU+16} \\ \hline
		\acs{BS} antenna pattern					& \(65^\circ\) \acs{HPBW} in azimuth and elevation \\ \hline
    	Number \acs{MS}	antennas \(n_t\)   & 1 \\ \hline
		\
		\acs{MS} antenna type							& Omni-directional vertically polarized \\ \hline
		\acs{SIMO} \acs{STTS} resolution	& \SI{1.08}{\milli\second} \\ \hline
		Covariance matrix window \(\tau\)	& \SI{10}{STTS} \\ \hline
		LTTS resolution										& \(\SI{0.66}{\second}\), see \cref{fig:track_view} \\ \hline

		Channel bandwidth									& \SI{18}{\mega\hertz}  \\ \hline
		Subcarrier bandwidth							& \SI{180}{\kilo\hertz} \\ \hline
		Number of \acs{OFDM} subcarriers \(N\)	& 100 \\
		
	\end{tabular}
	\caption{Measurement and data processing parameters.}
	\label{tab:sim_assump}
	\vspace{-0.35cm}
\end{table}
The main measurement and data processing parameters are listed in Table~\ref{tab:sim_assump}. For covariance matrix calculation according to \cref{eq:cov_measurements}, a window length of \(\tau=10\) \acp{STTS} is applied. Since no studies are available on reliable covariance matrix calculation \ac{w.r.t.} \(\tau\), the choice of \(\tau\) in our work is heuristic and requires further investigation. Next, we provide a classification of the measurement tracks, followed by a thorough analysis of the results.

\subsection{Track Classification}\label{sec:track_class}
After a thorough examination of each track with its serving \ac{BS} and corresponding surroundings shown in \cref{fig:meas_map}, we sort each measurement into one of the following classes:
\begin{itemize}
	\item \textbf{\acs{LoS} radial}: The track is approximately radial/perpendicular to the \ac{BS} and \ac{LoS} is available throughout the whole track.
	\item \textbf{\acs{LoS} tangential}: The track is approximately tangential \ac{w.r.t.} the position of the \ac{BS} with a \ac{LoS} condition over the complete track.
	\item \textbf{Far away}: The track is located far away (at least \SI{100}{\m}) from the \ac{BS} \ac{w.r.t.} the track distance of \(\approx \SI{40}{\meter}\). Only available in the open street scenario.
	\item \textbf{Uncorrelated}: Tracks with low correlation, with either \ac{NLoS} condition or obstructed \ac{LoS} condition\footnote{Obstructed \ac{LoS} means objects with a size similar to the wavelength between transmitter and receiver, e.g. three branches.}.
	\item \textbf{Other}: The curve shape of the spatial consistency does not follow the other classes.
\end{itemize}
Table~\ref{tab:track_classes} maps the measurement tracks from \cref{fig:meas_map} to the classes given above. Note that due to the highly dynamic transmission environment, e.g. transition from \ac{LoS} to \ac{NLoS} or vice versa, and obstacles that are not shown in the measurement map, e.g. parking cars, the above-mentioned classes can not cover all measurements. Therefore, we leave out evaluation of the measurements in the ``Other'' class for future work. We can see from the table that except for the ``Far away'' class, each class contains tracks from both ``Campus'' and ``Open Street'' scenario.

\begin{figure}
	\centering
		\includegraphics[width=\columnwidth]{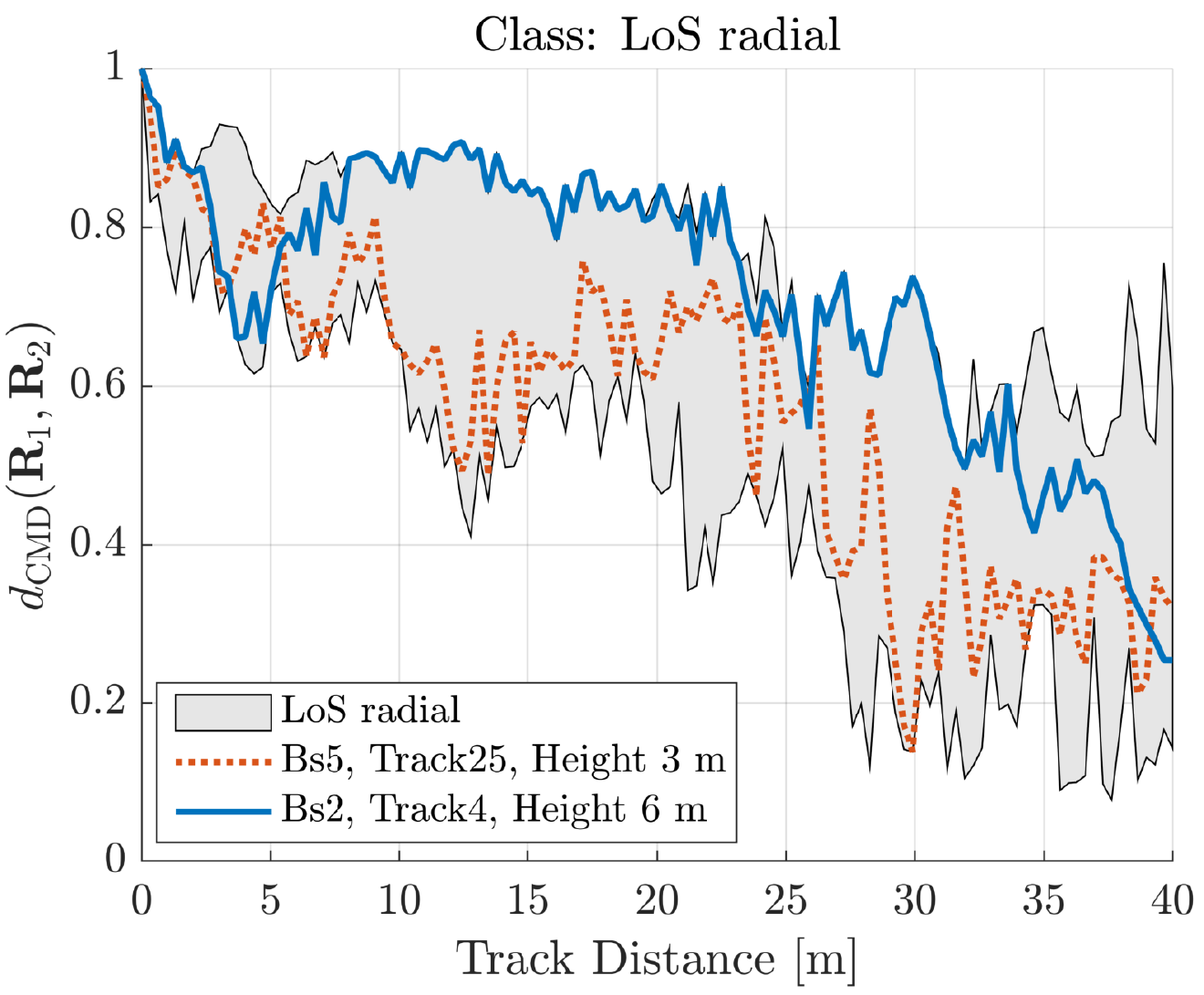}
		\caption{\ac{CMD} similarity over distance for the tracks in class ``\acs{LoS} radial'' according to \cref{tab:track_classes}. }
	\label{fig:LoS_radial}
	\vspace{-0.35cm}
\end{figure}

In order to study the behavior of the \ac{CMD} similarity \ac{w.r.t.} distance, we set the starting \ac{LTTS} of each measurement as user 1, while treating each of the remaining \acp{LTTS} as user 2 moving away from user 1. Therefore, the \(d_{\mathrm{CMD}}\) between two users can be calculated according to \cref{eq:cmd} for each track. \cref{fig:LoS_radial} shows the \ac{CMD} similarity of class ``\ac{LoS} radial'' over distance. For visualizing purpose, the area between the minimum and maximum \(d_{\mathrm{CMD}}\) over all tracks in the class at each \ac{LTTS} is shown as a gray cover plot. Otherwise, the many lines within the same figure would be hard to distinguish. Instead, two representative \(d_{\mathrm{CMD}}\) curves are given to show the dynamic range of the measurement tracks. We can see in \cref{fig:LoS_radial} that in this class the average \(d_{\mathrm{CMD}}\) drops steadily over distance but overall remains high, e.g. mostly above 0.5 up to \SI{20}{\m}. This high correlation can be explained by the \ac{LoS} condition and that most of the power is received by the \ac{LoS} path.

\begin{figure}
	\centering
		\includegraphics[width=\columnwidth]{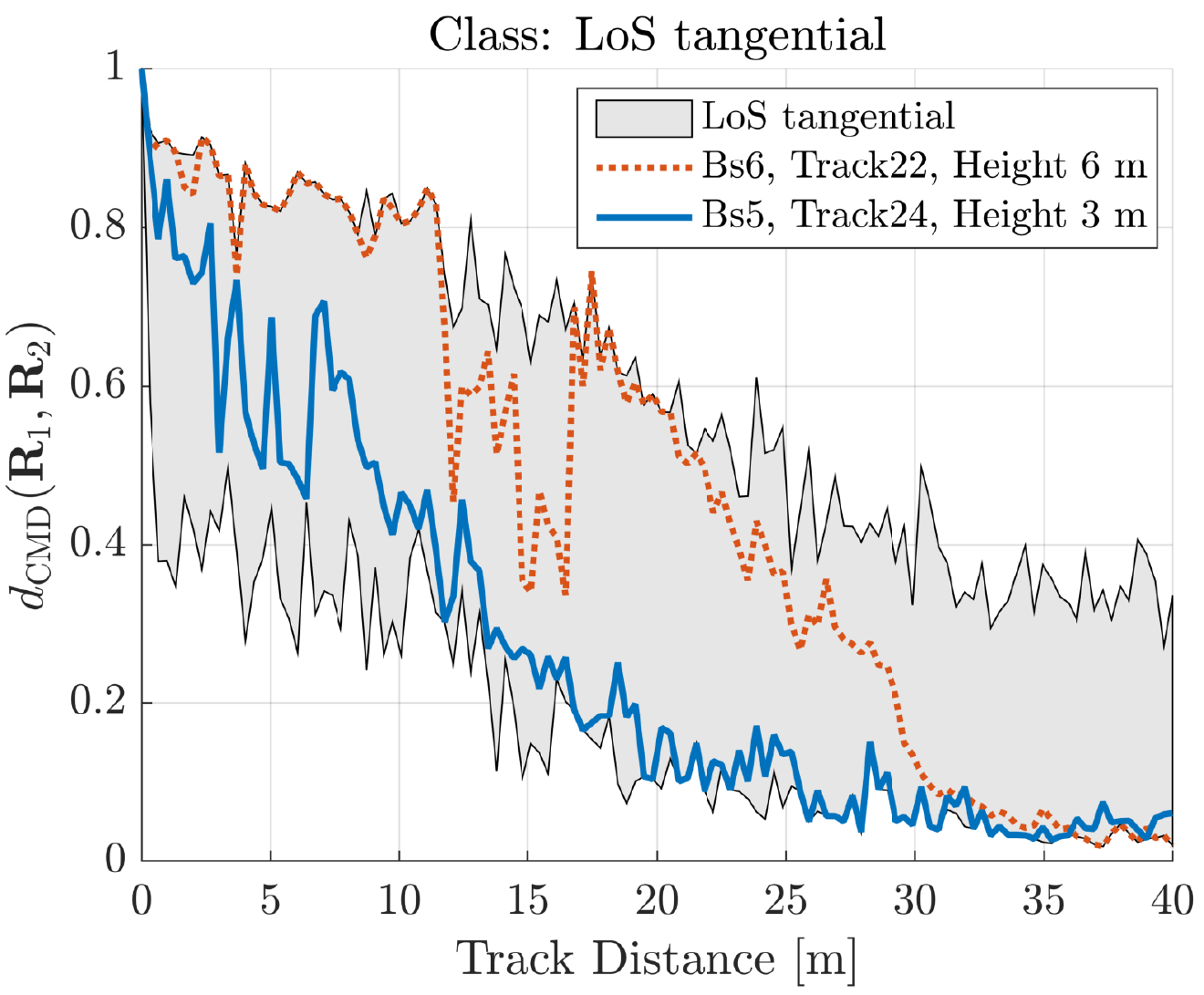}
		\caption{\ac{CMD} similarity over distance for the tracks in class ``\acs{LoS} tangential'' according to \cref{tab:track_classes}. }
	\label{fig:LoS_tangential}
	\vspace{-0.35cm}
\end{figure}

Similarly, \cref{fig:LoS_tangential} shows results for the ``\ac{LoS} tangential'' class. Here \(d_{\mathrm{CMD}}\) decreases almost proportionally with the distance down to 0.1. In comparison to the ``\ac{LoS} radial'' class, the ``\ac{LoS} tangential'' class decreases faster. This can be explained by the change in angles. In the ``\ac{LoS} radial'' class, where the tracks are perpendicular to the \ac{BS},  moving along tracks only changes the elevation angle of the \acp{MPC}, whereas in class ``\ac{LoS} tangential'', where the tracks are tangential to the \ac{BS}, moving along tracks changes both azimuth and elevation angles of the \acp{MPC}. The additional change in the azimuth angle of the \acp{MPC} causes the steeper decrease in \(d_{\mathrm{CMD}}\).

\begin{figure}
	\centering
		\includegraphics[width=\columnwidth]{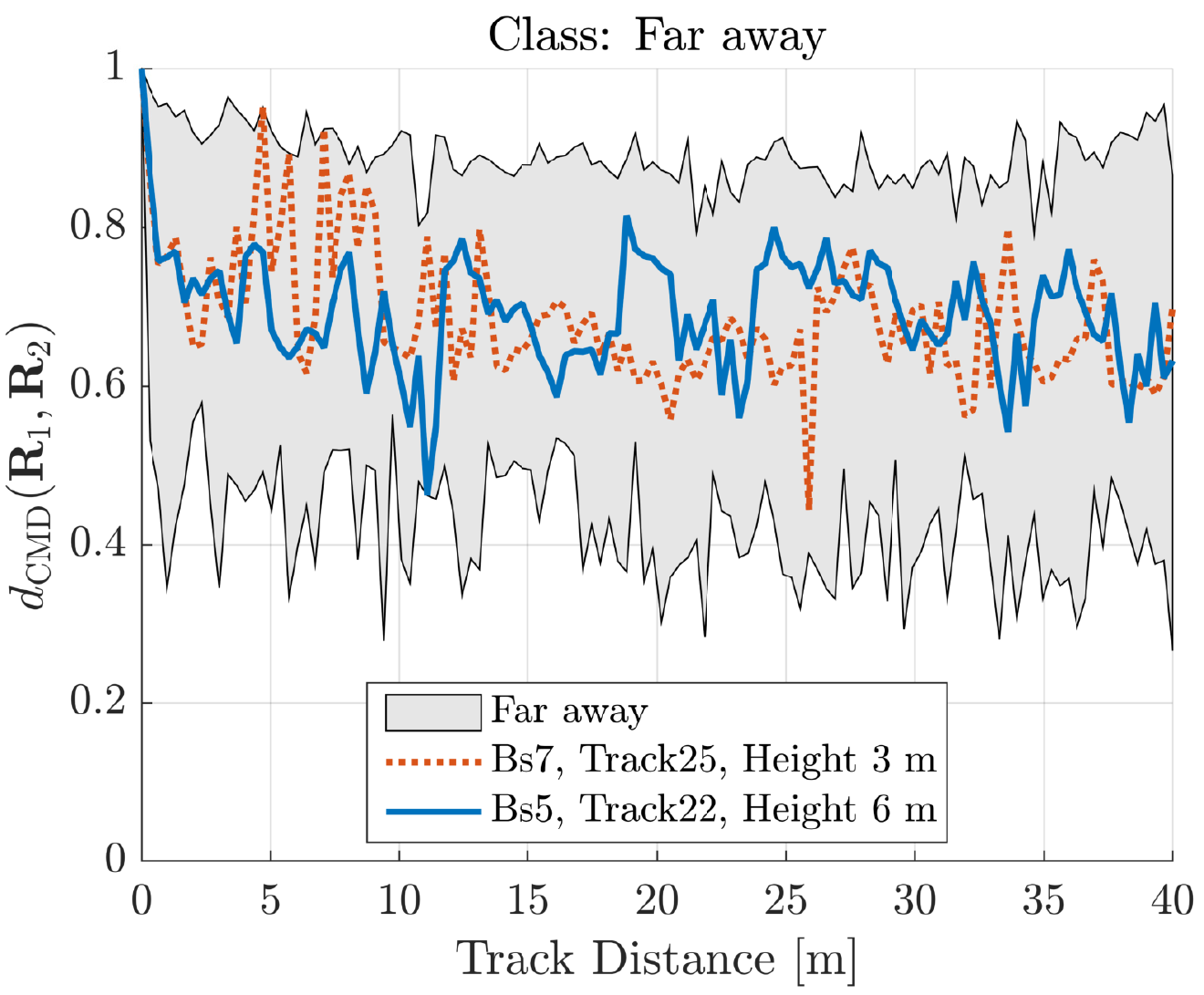}
		\caption{\ac{CMD} similarity over distance for the tracks in class ``Far away'' according to \cref{tab:track_classes}. }
	\label{fig:Far_Away}
	\vspace{-0.35cm}
\end{figure}

Next, the ``Far away'' class is shown in \cref{fig:Far_Away}. It is of interest to note that the \ac{CMD} similarity fluctuates over distance and no significant decrease is observed for tracks in this class. This is due to the fact that distance between users has a smaller impact on the channel for users located far away from the \ac{BS} than the ones which are closer to the \ac{BS}. To further elaborate, the change in channel coefficients is dependent on the ratio of the relative distance between users to the total distance from the \ac{BS} to the track. The smaller the ratio, the smaller the changes in \(d_{\mathrm{CMD}}\). 

\begin{figure}
	\centering
		\includegraphics[width=\columnwidth]{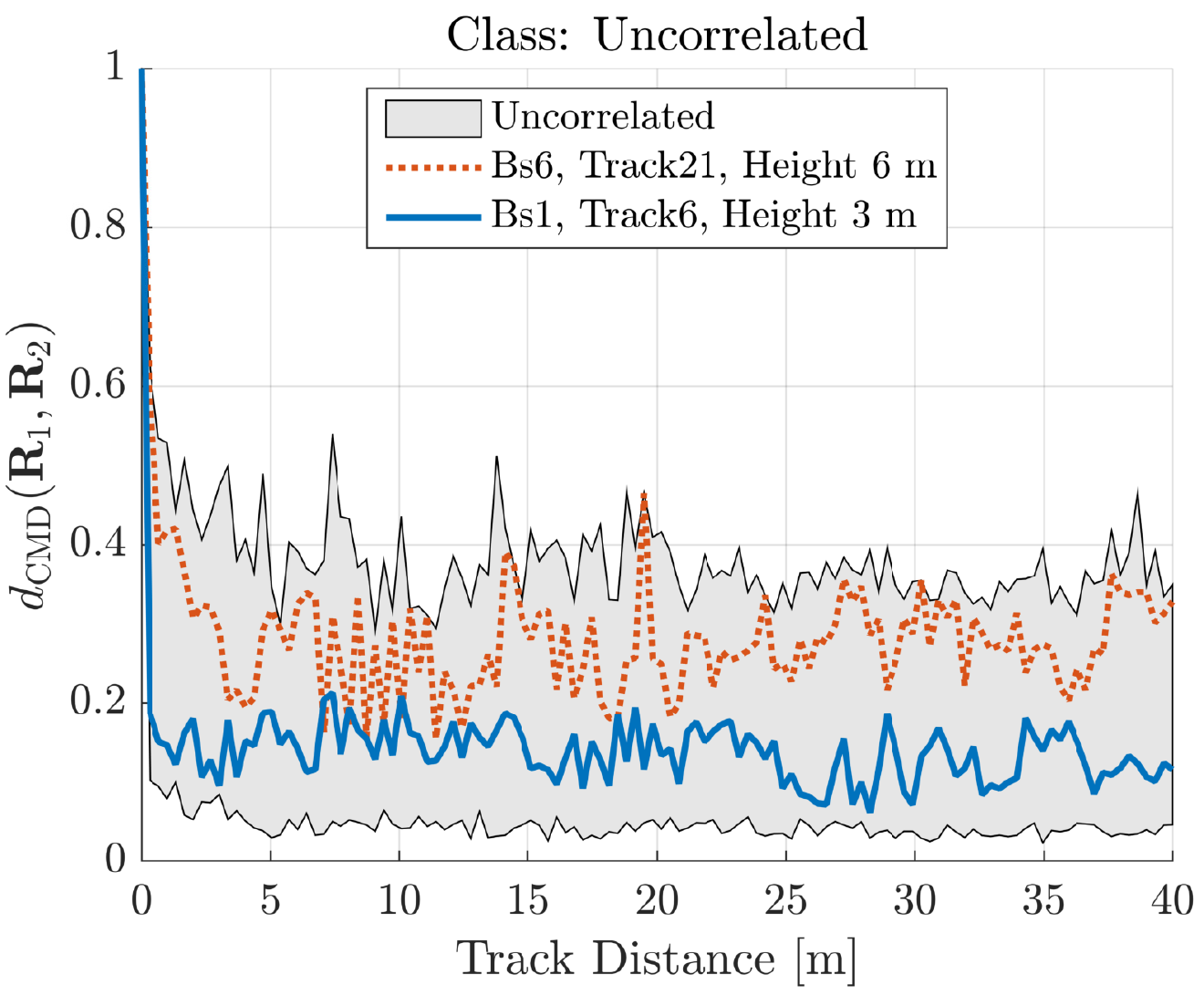}
		\caption{\ac{CMD} similarity over distance for the tracks in class ``Uncorrelated'' according to \cref{tab:track_classes}. }
	\label{fig:Uncorrelated}
	\vspace{-0.35cm}
\end{figure}

At last, \cref{fig:Uncorrelated} shows the ``Uncorrelated'' class. In this case, low correlation is observed in the measurements, since \(d_{\mathrm{CMD}}\) drops below 0.4 after the first few \acp{LTTS}. A cross-check with \cref{fig:meas_map} shows that most of these measurements are taken on tracks without direct \ac{LoS} to the serving \ac{BS}. This indicates that the angles of the received \acp{MPC} at the \ac{BS} change rapidly even by moving the transmitter for \SI{1}{\meter}. One explanation for this could be a fast change of scattering objects that results in non-continuous phase and amplitude jumps of \acp{MPC}. This is also called ``death'' and ``birth'', or lifetime, of scatterers. In our measurements this relates to scatterers observed by the \ac{BS} which was a \ac{UCA} deployed on a pole and therefore able to receive \acp{MPC} from all azimuth directions.

\subsection{Result Analysis}\label{sec:result_analysis}
These 4 classifications show that the spatial consistency depends less on the \acp{LSP} of a given scenario and more on the individual geometry, as both the ``Campus'' and ``Open Street'' scenarios appear in the same classes.
In general, the similarity of covariance matrices decreases over distance. However, depending on each classification, the similarity can have large variation locally, e.g. the similarity remains high over \SI{40}{\m} distances in the ``Far-Away'' category, whereas in the ``Uncorrelated'' category, the similarity is barely seen. 
Furthermore, the similarity of the covariance matrices is decreasing within a very short distance in many of the \ac{NLoS} scenarios. This indicates that clustering of users based on covariance matrix similarity can be difficult to achieve in \ac{NLoS} scenarios. 

The above described effects are not captured by the current \ac{3GPP} proposal of the spatial consistency feature where only a single dependency on the ``correlation distance'' and the distance between users is captured, see \cite{KDJT18}. Therefore, an extension of the spatial consistency feature in \cite{3GPP17-38901} is required.
\begin{table}
	\renewcommand{\tabcolsep}{1pt}
	\centering
	\begin{tabular}{C{1.3cm}|C{0.3cm}|C{4.8cm}}
		\hline
		Track Class 																	& \acs{BS} ids & Track ids \\ \hline \hline
		\multirow{4}{\linewidth}{\acs{LoS} radial} 	& 1  & 2 (\SI{6}{\meter})  \\ \cline{2-3} 
																									& 2  & 4 (\SI{6}{\meter}) \\ \cline{2-3} 
																									& 5  & 13, 25 \\ \cline{2-3}
																									& 6  & 24 (\SI{6}{\meter})\\ \hline
		\multirow{3}{\linewidth}{\acs{LoS} tangential} 	& 1  &  4 (\SI{3}{\meter})\\ \cline{2-3}
																									& 2  &  3 \\ \cline{2-3}
																									& 3  &  4 (\SI{6}{\meter})\\ \cline{2-3}
																									& 5  & 23, 24 \\ \cline{2-3} 
																									& 6  &  14, 15 (\SI{3}{\meter}), 22\\ \cline{2-3}
																									& 7  & 17 (\SI{6}{\meter}), 18 (\SI{3}{\meter}), 19 (\SI{3}{\meter}), 20 (\SI{3}{\meter}) \\ \hline
		\multirow{3}{\linewidth}{Far away}						& 5  & 15 (\SI{3}{\meter}), 16, 18 (\SI{3}{\meter}), 19, 20 (\SI{3}{\meter}), 21, 22 \\ \cline{2-3} 
																									& 6  & 17 (\SI{6}{\meter}), 18 (\SI{6}{\meter}), 19  (\SI{6}{\meter}) \\ \cline{2-3} 
																									& 7  & 13 (\SI{6}{\meter}), 15 (\SI{3}{\meter}),23 (\SI{6}{\meter}), 25 (\SI{3}{\meter}) \\ \hline
		\multirow{4}{\linewidth}{Un-correlated Tracks}	
																									& 1 & 5 (\SI{6}{\meter}), 6, 7, 9, 10, 11, 12 \\ \cline{2-3}
																									& 2 & 1 (\SI{6}{\meter}), 2 (\SI{6}{\meter}), 5, 6, 7, 8, 10 (\SI{6}{\meter}) \\ \cline{2-3}
																									& 3 & 2, 5 (\SI{6}{\meter}), 9, 10 (\SI{3}{\meter}), 11, 12 \\ \cline{2-3}
																									& 5 & 26 (\SI{3}{\meter}) \\ \cline{2-3}
																									& 6 & 21 (\SI{6}{\meter}) \\ \hline
	\end{tabular}
	\caption{Mapping of measurement tracks to track-classes}
	\label{tab:track_classes}
	\vspace{-0.35cm}
\end{table}

